\documentclass[fleqn,10pt]{wlscirep}
\usepackage[utf8]{inputenc}
\usepackage[T1]{fontenc}
\usepackage{changepage}
\usepackage{multirow}
\usepackage{makecell}
\usepackage{color}
\usepackage{comment}

\usepackage{floatrow}
\usepackage{graphicx}
\usepackage[label font=bf,labelformat=simple]{subfig}
\usepackage{caption}
\title{Haplotype-resolved \emph{de novo} assembly with phased assembly graphs}

\author[1,2]{Haoyu Cheng}
\author[3]{Gregory T Concepcion}
\author[1,2]{Xiaowen Feng}
\author[4]{Haowen Zhang}
\author[1,2,*]{Heng Li}
\affil[1]{Department of Data Sciences, Dana-Farber Cancer Institute, Boston, MA, USA}
\affil[2]{Department of Biomedical Informatics, Harvard Medical School, Boston, MA, USA}
\affil[3]{Pacific Biosciences, Menlo Park, CA, USA}
\affil[4]{College of Computing, Georgia Institute of Technology, Atlanta, GA, USA}

\affil[*]{To whom correspondence should be addressed: hli@jimmy.harvard.edu}

\begin{abstract}
Haplotype-resolved \emph{de novo} assembly is the ultimate solution to the
study of sequence variations in a genome. However, existing algorithms either
collapse heterozygous alleles into one consensus copy or fail to cleanly
separate the haplotypes to produce high-quality phased assemblies. Here we
describe hifiasm, a new \emph{de novo} assembler that takes advantage of
long high-fidelity sequence reads to faithfully represent the haplotype
information in a phased assembly graph. Unlike other graph-based assemblers
that only aim to maintain the contiguity of one haplotype, hifiasm strives to
preserve the contiguity of all haplotypes. This feature enables the development
of a graph trio binning algorithm that greatly advances over standard trio
binning. On three human and five non-human datasets, including California
redwood with a $\sim$30-gigabase hexaploid genome, we show that hifiasm
frequently delivers better assemblies than existing tools and consistently
outperforms others on haplotype-resolved assembly.
\end{abstract}

\begin{document}

\flushbottom
\maketitle
\thispagestyle{empty}

\section*{Introduction}

\emph{De novo} genome assembly is the most comprehensive method that provides
unbiased insight to DNA sequences. With the rapid advances in long-read
sequencing technologies such as Pacfic Biosciences (PacBio) and Oxford Nanopore
(ONT), many long-read assemblers have been developed to tackle this essential
computational problem. Most of them
\cite{chin2013nonhybrid,Berlin:2015xy,li2016minimap,koren2017canu,kolmogorov2019assembly,chin2019human,ruan2020fast,shafin2020nanopore,Chen2020.02.01.930107}
collapse different homologous haplotypes into a consensus representation with
heterozygous alleles frequently switching in the consensus. This approach works
well for inbred samples that are nearly homozygous but necessarily misses half
of the genetic information in a diploid genome. To solve this problem, Falcon-Unzip
\cite{chin2016phased} recovers heterozygous alleles by ``unzipping'' them in an
initial collapsed assembly. It produces a pair of assemblies, one primary
assembly representing a mosaic of homologous haplotypes, and one alternate
assembly composed of short haplotype-specific contigs (haplotigs) for alleles
absent from the primary assembly. The alternate assembly is often fragmented
and does not represent a complete haplotype, making it less useful in practice.
In addition, starting from a collapsed assembly, Falcon-Unzip may not recover
highly heterozygous regions not properly collapsed in the initial assembly.
Trio binning \cite{koren2018novo} addresses these issues by globally
partitioning long reads upfront with parental short reads and then performing
two separate assemblies on the partitioned reads. This strategy works well for
samples with high heterozygosity, but for a human sample sequenced with noisy
long reads, it only produces fragmented assemblies with $\sim$1.2~Mb contigs.

A great challenge to the assembly of heterozygous samples is caused by the
5--15\% sequencing error rate of older long reads. With this high error rate,
it is difficult to distinguish errors from heterozygotes occurring at a rate of
<0.1\% in humans. The recent availability of high-fidelity (HiFi) reads
\cite{wenger2019accurate} produced by PacBio has changed the equation.
Generated from the consensus of multiple sequences of the same DNA molecule,
HiFi reads have a much lower error rate of <1\%. With HiFi, standard trio
binning can produce contigs of $~$17~Mb \cite{wenger2019accurate}. Recent work
relying on Hi-C or Strand-seq read binning
\cite{garg2019efficient,porubsky2019fully} can achieve better contiguity and
phasing accuracy. These pre-binning algorithms all use short k-mers or short
reads to partition HiFi reads. They may not identify haplotype-specific markers
in complex regions and result in wrong read partitions which will negatively affect the
assembly as we will show later. In addition, both Hi-C and Strand-seq
binning start with a collapsed assembly and have the same issues as
Falcon-Unzip.

In 2012, we reasoned \cite{Li:2012fk} that a perfectly constructed unitig graph
with read information is a lossless representation of single-end reads. Because
this graph is lossless, we can compress input reads into a unitig graph and
perform phasing later. This should maximize the power of long HiFi reads.
Developed in parallel to our work, HiCanu \cite{nurk2020hicanu} follows a
similar rationale and can produce Falcon-Unzip-style primary/alternate
assemblies better than other assemblers especially around segmental
duplications. However, HiCanu only tries to keep the contiguity of one parental
haplotype and often breaks the contiguity of the other haplotype. When we
separate parental haplotypes, these break points will lead to fragmented
haplotype-resolved assemblies. HiCanu is not making use of the full power of
HiFi reads.

In this article we present hifiasm, a new assembler for HiFi reads that
generates a well connected assembly graph and produces better assemblies in
practice. We will first give an overview of the hifiasm algorithm, compare it
to other assemblers for partially phased assemblies and then explain and
evaluate the haplotype-resolved assembly algorithm used by hifiasm.

\begin{figure}[!t]
\centering
\includegraphics[width=0.72\textwidth]{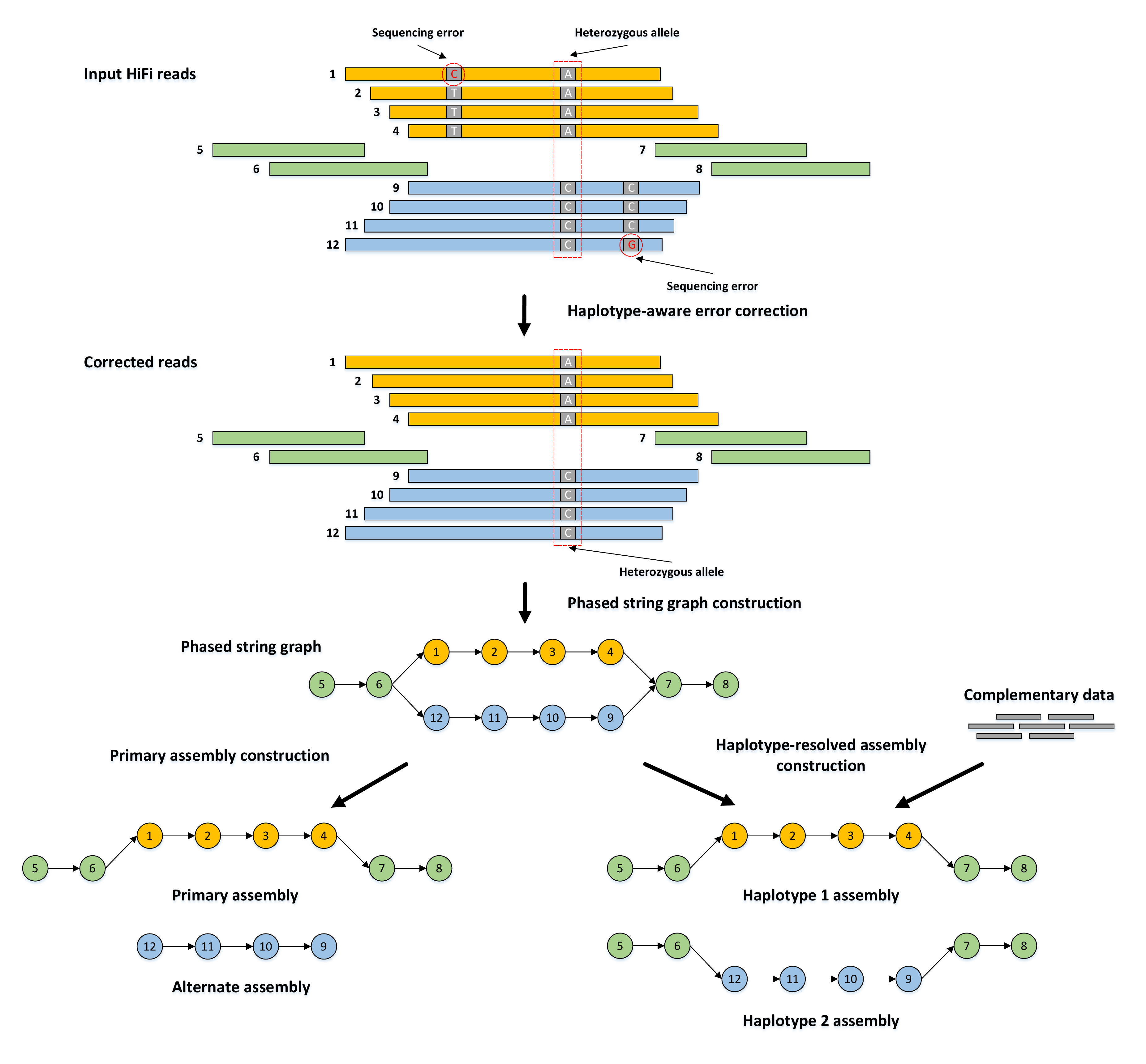}
\caption{{\bf Outline of the hifiasm algorithm.} Reads in orange and blue
represent the reads with heterozygous alleles carrying local phasing
information, while reads in green come from the homozygous regions without any
heterozygous alleles.  In phased string graph, a vertex corresponds to the
HiFi read with same ID, and an edge between two vertices indicates that their
corresponding reads are overlapped with each other. Hifiasm first performs
haplotype-aware error correction to correct sequence errors but keep
heterozygous alleles, and then builds phased assembly graph with local phasing
information from the corrected reads. Only the reads coming from the same
haplotype are connected in the phased assembly graph. With complementary data
providing global phasing information, hifiasm generates a completely phased
assembly for each haplotype from the graph. Hifiasm also can generate unphased
primary assembly only with HiFi reads. This unphased primary assembly
represents phased blocks (regions) that are resolvable with HiFi reads, but
does not preserve phasing information between two phased blocks.}
\label{figf1}
\end{figure}

\section*{Results}

\subsection*{Overview of the hifiasm algorithm}

The first few steps of hifiasm broadly resemble the workflow of early long-read
assemblers \cite{chin2013nonhybrid,Berlin:2015xy} (Fig.~\ref{figf1}). Hifiasm
performs all-vs-all read overlap alignment and then corrects sequencing errors.
Given a target read to be corrected, hifiasm inspects the alignment of reads
overlapping with the target read. A position on the target read is said to be
informative if there are two types of A/C/G/T bases (gaps ignored) at the
position in the alignment and each type is supported by at least three reads.
A read overlapping with the target read is inconsistent with the target if
there are informative positions in the overlap and the read is not identical to
the target read across all these positions; accordingly, the overlap between
this and the target read is inconsistent. Inconsistent reads theoretically
originate from a haplotype different from the target read. Hifiasm only uses
consistent reads to correct the target read.

Hifiasm performs three rounds of error correction by default. It then does
overlap alignment again and builds a string graph \cite{myers2005fragment}
where a vertex is an oriented read and an edge is a consistent overlap. After
transitive reduction, a pair of heterozygous alleles will be represented by a
``bubble'' in the string graph (Fig.~\ref{figf1}). No information is lost.
If there are no additional data, hifiasm arbitrarily selects
one side of each bubble and outputs a primary assembly similar to Falcon-Unzip
and HiCanu. For a heterozygous genome, the primary assembly achieved at this step
may still contain haplotigs from more than one homologous haplotypes. With
HiCanu, we have to post-process the assembly with purge\_dups
\cite{guan2019identifying} to remove these redundant haplotigs. Hifiasm
natively implements a variant of the purge\_dups algorithm to identify and
remove such haplotigs. This simplifies the assembly pipeline.

If parents of the sample are also sequenced, hifiasm can use k-mer trio binning
\cite{koren2018novo} to label corrected reads in the string graph. In this
case, hifiasm effectively discards the maternal unitigs to generate the paternal assembly,
and vice versa. This graph-based trio binning may go through regions
heterozygous in all three samples in the trio and is more robust to the
mislabeling of reads. We will explain the advantage of hifiasm binning in a
later section.

\begin{table}[!tb]
\captionsetup{singlelinecheck = false, justification=justified}
\footnotesize
\caption{Statistics of non-human assemblies}
{
\begin{tabular*}{\textwidth}{@{\extracolsep{\fill}} llrrrrcc}

\hline
\multirow{2}{*}{Dataset} & \multirow{2}{*}{Assembler} & \makecell[c]{Size} & \makecell[c]{N50} & \makecell[c]{NG50} & \makecell[c]{Alternate} & \multicolumn{2}{c}{\makecell[c]{Completeness (asmgene or BUSCO)}}\\
\cline{7-8}
                         &                            & (Gb)               & (Mb)              & (Mb)               & size (Gb)               & Complete (\%)          & Duplicated (\%) \\

\hline
\multirow{4}{*}{\makecell{\textit{M. musculus} \\(25$\times$)}}
& hifiasm        & 2.606 & 21.1 & 20.6 & 0.087 & 99.72 & 0.23\\
& HiCanu         & 2.604 & 16.4 & 15.9 & 0.033 & 99.70 & 0.23\\
& Peregrine      & 2.578 & 17.9 & 17.0 & 0.029 & 99.56 & 0.21\\
& Falcon         & 2.559 & 19.3 & 16.7 & 0.025 & 99.49 & 0.14\\

\hline
\multirow{4}{*}{\makecell{\textit{Z. mays} \\(22$\times$)}}
& hifiasm        & 2.190 & 36.7 & 36.7 & 0.106 & 99.84 & 0.18\\
& HiCanu         & 2.183 & 14.5 & 14.5 & 0.284 & 99.85 & 0.48\\
& Peregrine      & 2.206 & 1.8  & 1.9  & 0.117 & 99.79 & 0.27\\
& Falcon         & 2.132 & 9.5  & 9.3  & 0.016 & 99.77 & 0.17\\

\hline
\multirow{5}{*}{\makecell{\textit{F. $\times$ ananassa} \\(36$\times$)}}
& hifiasm (purge)& 0.826 & 17.8 & 17.8 & 0.473 & 98.51 & 93.06 \\
& HiCanu         & 1.214 & 8.7  & 14.8 & 0.098 & 97.96 & 92.87 \\
& HiCanu (purge) & 0.381 & 10.9 &      & 0.931 & 93.56 & 41.26 \\
& Peregrine      & 0.930 & 5.5  & 6.7  & 0.260 & 98.33 & 91.70 \\
& Falcon         & 0.971 & 5.4  & 7.3  & 0.213 & 98.27 & 92.81 \\

\hline
\multirow{2}{*}{\makecell{\textit{R. muscosa} \\($\sim$29$\times$)}}
& hifiasm (purge)& 9.535 & 9.3 & & 7.588 & 66.81 & 1.59 \\
& Peregrine      & 9.415 & 0.9 & & 2.936 & 66.84 & 1.72 \\

\hline
\multirow{2}{*}{\makecell{\emph{S. sempervirens} \\($\sim$28$\times$)}}
& hifiasm (purge)& 35.624 & 5.4 & & 16.306 & 61.82 & 39.85 \\
& Peregrine      & 35.662 & 0.8 & &        & 63.20 & 35.93 \\

\hline
\end{tabular*}
}

\begin{flushleft} \footnotesize{
HiCanu (purge) applies Purge\_dups to a HiCanu assembly. Hifiasm (purge) enables the built-in Purge\_dups equivalent strategy.
The N50/NG50 of an assembly is defined as the sequence length of the shortest contig at 50\% of
the total assembly/genome size. To calculate the NG50, a genome size of 2730.9
Mb (AC:GCF\_000001635.20), 2182.1 Mb (AC:GCA\_902167145.1)
and 813.4 Mb \cite{edger2019origin} is used for \emph{M.
musculus}, \emph{Z.  mays} and \emph{F. $\times$ ananassa}, respectively. The
genome size is unknown for \emph{R. muscosa} and \emph{S.  sempervirens}. The
NG50 of HiCanu (Purge) is not available since its size is less than 50\% of the
genome size. ``Alternate size'' is the total length of the alternate assembly.
The reference-based asmgene method \cite{li2018minimap2} is used to evaluate the gene completeness of
\emph{M. musculus} and \emph{Z. mays} which have high-quality
reference genomes. For these two samples, ``Complete'' gives the percentage of single-copy genes in
the reference genome (one unique mapping at $\ge$97\% identity) that are
mapped at $\ge$97\% identity to the assembly; ``Duplicated'' gives the percentage
of reference single-copy genes that become multi-copy in the assembly. 
The BUSCO embryophyta dataset is used to evaluate the gene completeness of \emph{F.
$\times$ ananassa} and \emph{S. sempervirens}; the tetrapoda dataset is used
for \emph{R. muscosa}.}
\end{flushleft} \label{table1}

\end{table}

\subsection*{Assembling homozygous non-human genomes}

We first evaluated hifiasm v0.7 along with Falcon-Unzip v1.8.1 \cite{chin2016phased},
Peregrine v0.1.6.1 \cite{chin2019human} and HiCanu v2.0 \cite{nurk2020hicanu} on two inbred
samples \cite{Hon2020.05.04.077180} including the C57/BL6J strain of \emph{M.
musculus} (mouse) and the B73 strain of \emph{Z. mays} (maize). All assemblers
produced long contigs for mouse (Table~\ref{table1}). To evaluate how often
assemblers collapse paralogous regions and produce a misassembly, we mapped HiFi
reads to each assembly, extracted apparently heterozygous SNPs at high coverage
and clustered them into longer regions (Online Methods).  These regions
correspond to collapsed misassemblies. We identified 5 such misassemblies in
the HiCanu assembly, 17 in hifiasm and more than 100 in both Falcon and
Peregrine. HiCanu is the best at this metric although its contig N50 is the
shortest.

For the repeat-rich maize genome, hifiasm and HiCanu generated longer contigs
and again produced much fewer collapsed misassemblies (2 for hifiasm and 3 for
HiCanu, versus more than 70 misassemblies for Falcon and Peregrine). Hifiasm and
HiCanu perform better presumably because they can more effectively resolve
repeats by requiring near perfect overlap \cite{nurk2020hicanu}.

\subsection*{Assembling heterozygous non-human genomes}

Since most practical samples are heterozygous, we next evaluated the assemblers
on three heterozygous datasets from \emph{F. $\times$ ananassa} (garden
strawberry), \emph{R. muscosa} (mountain yellow-legged frog) and
\emph{S. sempervirens} (California redwood). These samples are more
challenging to assemble. \emph{F. $\times$ ananassa} has an allopolyploid
genome estimated to be 813.4 Mb in size \cite{edger2019origin}. All assemblers
achieve a total assembly of $\sim$1.2~Gb, including both primary and alternate
contigs. However, they resolve the primary assembly differently. Hifiasm results in a
primary assembly of similar size to the published genome. BUSCO
\cite{simao2015busco} regards most single-copy genes to be duplicated, consistent
with the previous observation \cite{edger2019origin}. HiCanu assigns most
contigs to the primary. Applying Purge\_dups \cite{guan2019identifying}
overcompresses the assembly and reduces the BUSCO completeness by 5\%.
Falcon-Unzip and Peregrine are somewhat between hifiasm and HiCanu. The varying
primary assembly sizes highlight the difficulty in assembling polyploid
genomes. On the other hand, all HiFi assemblies here have much longer contig
N50 than the published assembly (>5 Mb vs 580 kb). HiFi enables better
assembly.

\emph{R. muscosa} is hard to assemble for its large genome size. We failed to
run Falcon-Unzip and HiCanu for this sample using their released versions.
Both hifiasm and Peregrine were successful. The N50 of the hifiasm assembly is ten
times as long.

\emph{S. sempervirens} poses an even greater challenge to assembly with a much
larger hexaploid genome. Hifiasm took 875 Gb reads as input and produced a 35.6
Gb assembly in 2.5 days over 80 CPU threads using $\sim$700 GB memory at the
peak. The flow cytometric estimate of the full hexaploid genome is 62.8 Gb in
size \cite{Hizume_2001}. Our assembly is about half of that. Peregrine achieved an
35.6 Gb assembly as well. Its BUSCO score is 1.4\% better than the hifiasm
assembly. However, BUSCO completeness may not be reliable. Peregrine also took 10 days on a computer cluster. It runs slower
and its assembly is more fragmented. Hifiasm overall performs better on large
genomes.

\begin{table}[!tb]
\captionsetup{singlelinecheck = false, justification=justified}
\footnotesize
\caption{Statistics of human primary assemblies}
{
\begin{tabular*}{\textwidth}{@{\extracolsep{\fill}} llcccccccc}

\hline
\multirow{2}{*}{Dataset} & \multirow{2}{*}{Assembly} & \makecell[c]{Size} & NG50 & NGA50 & \multirow{2}{*}{QV} & \makecell[c]{Multi-copy genes} & \makecell[c]{Resolved}
	& \multicolumn{2}{c}{\makecell[c]{Gene completeness (asmgene)}}\\
\cline{9-10}
& & (Gb) & (Mb) & (Mb) &  & retained (\%) & BACs (\%) & Complete (\%) & Duplicated (\%) \\

\hline
\multirow{4}{*}{\makecell{CHM13 \\ (HiFi 32$\times$ \\ ONT 120$\times$)}}
& hifiasm        & 3.043 & 88.1 & 65.4 & 54.3 & 76.9 & 95.3 & 99.14 & 0.28 \\
& HiCanu         & 3.047 & 76.3 & 59.4 & 53.9 & 76.7 & 96.5 & 99.13 & 0.33 \\
& Peregrine      & 2.990 & 36.5 & 33.2 & 43.8 & 41.4 & 38.4 & 98.84 & 0.26 \\
& Falcon         & 2.862 & 26.3 & 23.8 & 50.1 & 24.6 & 33.1 & 98.62 & 0.11 \\
& Canu (ONT)     & 2.992 & 74.1 & 60.5 & 26.6 & 61.6 & 92.1 & 97.79 & 0.27 \\

\hline
\multirow{4}{*}{\makecell{HG00733 \\ (HiFi 33$\times$ \\ ONT 50$\times$)}}
& hifiasm (purge)& 3.039 & 70.0 & 56.8 & 49.8 & 67.3 & 83.2 & 99.09 & 0.31 \\
& HiCanu (purge) & 2.932 & 35.2 & 31.6 & 50.7 & 62.4 & 73.7 & 97.76 & 0.33 \\
& Peregrine      & 3.035 & 30.1 & 30.1 & 40.5 & 37.2 & 38.5 & 98.70 & 0.31 \\
& Falcon         & 2.861 & 24.4 & 23.2 & 46.3 & 33.6 & 38.0 & 96.51 & 0.15 \\
& Canu (ONT)     & 2.834 & 40.5 & 35.1 & 22.7 & 22.5 & 69.3 & 91.26 & 0.14 \\

\hline
\multirow{4}{*}{\makecell{HG002 \\ (HiFi 36$\times$ \\ ONT 80$\times$)}}
& hifiasm (purge)& 3.063 & 98.7 & 65.4 & 51.4 & 74.8 & & 99.31 & 0.35 \\
& HiCanu (purge) & 3.000 & 44.7 & 35.9 & 52.1 & 67.1 & & 98.97 & 0.23 \\
& Peregrine      & 3.081 & 33.4 & 32.5 & 41.3 & 42.5 & & 99.14 & 0.36 \\
& Falcon         & 2.955 & 30.4 & 29.0 & 46.7 & 36.6 & & 99.00 & 0.20 \\
& Canu (ONT)     & 2.831 & 32.3 & 30.5 & 21.9 & 19.6 & & 88.94 & 0.21 \\

\hline
\end{tabular*}
}

\begin{flushleft} \footnotesize{
Assemblies of ONT reads were obtained from other publications
\cite{miga2019telomere,shafin2020nanopore}; other assemblers use HiFi only.
HiCanu and hifiasm were run without duplication purging for the homozygous CHM13 cell line,
and run with purging for the heterozygous HG00733 and HG002 cell lines.
The NGA50 of an assembly is defined
as the length of the correctly aligned block at 50\% of the total reference
genome size which is assumed to be 3.1 Gb. It is calculated based on the
minigraph \cite{li2020design} contig-to-GRCh38 alignment. The ``QV'' (quality
value) of an assembly equals the Phred-scaled contig base error rate measured
by comparing 31-mers in contigs to 31-mers in short reads from the same sample.
Percent ``multi-copy genes retained'' is reported by asmgene (Online Methods).
It is the percentage of multi-copy genes in GRCh38 (multiple mapping positions
at $\ge$99\% sequence identity) that remain multi-copy in the assembly. A BAC is
resolved if 99.5\% of its bases can be mapped the assembly. There are 341 BACs
for CHM13 and 179 BACs for HG00733. HG002 does not have BAC data. The low gene
completeness numbers for ONT assemblies are caused by their low QV.}
\end{flushleft} \label{table2}

\end{table}

\subsection*{Primary assembly of human genomes}

We next evaluated hifiasm and other assemblers on three human datasets
(Table~\ref{table1}). We introduced two new metrics, ``multi-copy genes
retained'' and ``resolved BACs'' to evaluate how assemblers resolve difficult
genomic regions such as long segmental duplications. If an assembler breaks
contigs at such regions or misassemblies the regions, the resulting assembly
will lose multi-copy genes and/or leads to fragmented BAC-to-contig alignment.
We also used NGA50 to measure misassemblies. However, alignment cannot go
through assembly gaps in GRCh38 or long insertions/deletions. NGA50
is often an underestimate.

CHM13 is a homozygous cell line, similar to \emph{M. musculus} and \emph{Z.
mays}. Hifiasm delivers a more contiguous assembly. HiCanu resolves a few more
BACs, but it produced slightly more collapsed misassemblies (26 for HiCanu vs 20
for hifiasm). The two assemblers are broadly comparable. Both of them are
better than Peregrine, Falcon and the Canu ONT assembly \cite{miga2019telomere}
on all metrics by a large margin.

For heterozygous human samples HG00733 and HG002, HiCanu produced primary
assemblies of >3.5 Gb in size with several hundred megabases of heterozygous
regions represented twice. We had to run Purge\_dups
\cite{guan2019identifying} to remove these falsely duplicated regions in the
primary assembly. We tried a few Purge\_dups settings, including the default,
and chose the one that gave the best primary assembly. Hifiasm can identify and
remove falsely duplicated regions by inspecting inconsistent read overlaps
between them. Peregrine, Falcon and Shasta collapse most heterozygous regions
during assembly. They do not need additional tools like Purge\_dups, either.

For the two heterozygous samples, hifiasm and HiCanu consistently outperform
other assemblers. The hifiasm assembly is more complete and resolves more
difficult regions than HiCanu. This difference probably has more to do with the
duplicate purging algorithm than with the capability of the assembler.
Nonetheless, this observation suggests it is easier to produce a high-quality
primary assembly with hifiasm.

On running time, hifiasm takes 7--9 wall-clock hours over 48 threads. The peak
memory is below 150 GB. Peregrine is about twice as fast for human assembly but
uses more memory. HiCanu is about eight times as slow as hifiasm using the same
machine. Falcon is the slowest.

\begin{figure}[t]
\centering
\includegraphics[width=0.80\textwidth]{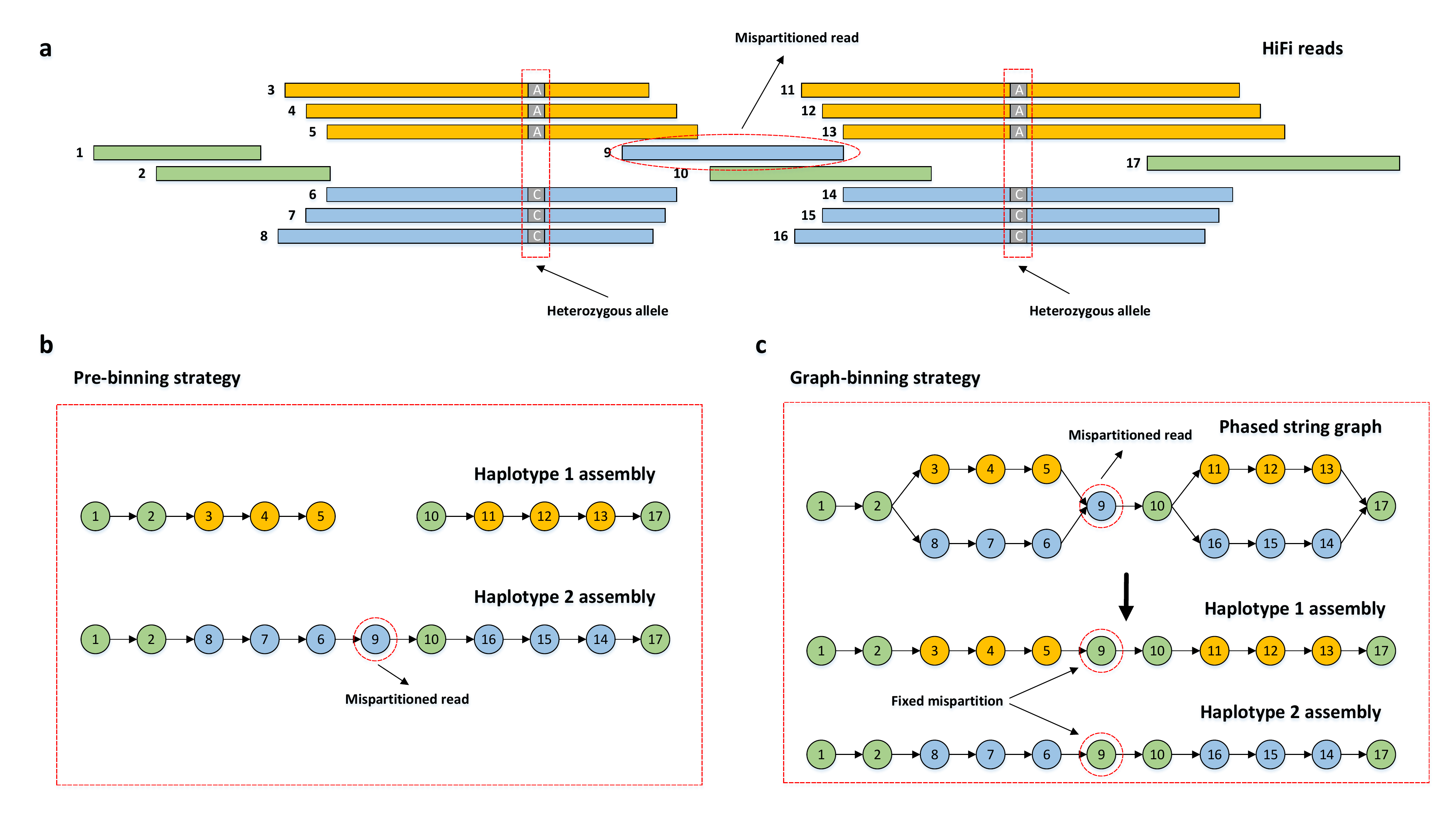}
\caption{{\bf Effect of false read binning.} \textbf{(a)} a set of reads with
global phasing information provided by the complementary data. Reads in orange
and reads in blue are specifically partitioned into haplotype 1 and haplotype
2, respectively. The remaining reads in green are partitioned into both
haplotypes. Read 9 without heterozygous alleles is mispartitioned into
haplotype 2, instead of to both haplotypes.  \textbf{(b)} A pre-binning
assembly produced by current methods which independently assemble two
haplotypes. Haplotype 1 is broken into two contigs due to the mispartition of
read 9. \textbf{(c)} Hifiasm fixes the mispartition by the local phasing
information in the phased assembly graph. It is able to identify that read 9
does not have heterozygous alleles, so that read 9 should be partitioned into
both haplotypes.} \label{figf2}
\end{figure}

\subsection*{Improving haplotype-resolved assembly}

A major issue with trio binning is that a fraction of heterozygous reads cannot
be unambiguously partitioned to parental haplotypes: if both parents are
heterozygous at a locus, a child read will harbor no informative k-mers and
cannot be uniquely assigned to a parental haplotype; if, say, the father is
heterozygous at a locus and the mother is homozygous, reads from the maternal
haplotype cannot be partitioned, either. With standard trio binning,
heterozygous reads that cannot be partitioned will be used in both parental
assemblies. As a result, both alleles may be present in one haplotype assembly
and lead to false duplications.  Standard trio binning is unable to cleanly
separate the two parental haplotypes.

Hifiasm draws power from HiFi read phasing in addition to trio binning. It does
not partition reads upfront; it only labels reads in the string graph. In a
long bubble representing a pair of heterozygous alleles, hifiasm may correctly
phase it even if only a small fraction of reads are correctly labeled. This way
hifiasm also rarely puts two alleles in one haplotype assembly.

Hi-C or Strand-seq based phasing \cite{garg2019efficient,porubsky2019fully} can
unambiguously phase most heterozygous reads and are naturally immune to false
duplications. They however suffer from another issue shared by standard trio
phasing: reads assigned to a wrong parental haplotype may break contigs
(Fig.~\ref{figf2}). By considering HiFi read phasing and the structure of the
assembly graph, hifiasm may be able to identify and fix such binning errors.

\begin{figure}[tb]
\centering
\includegraphics[width=0.45\textwidth]{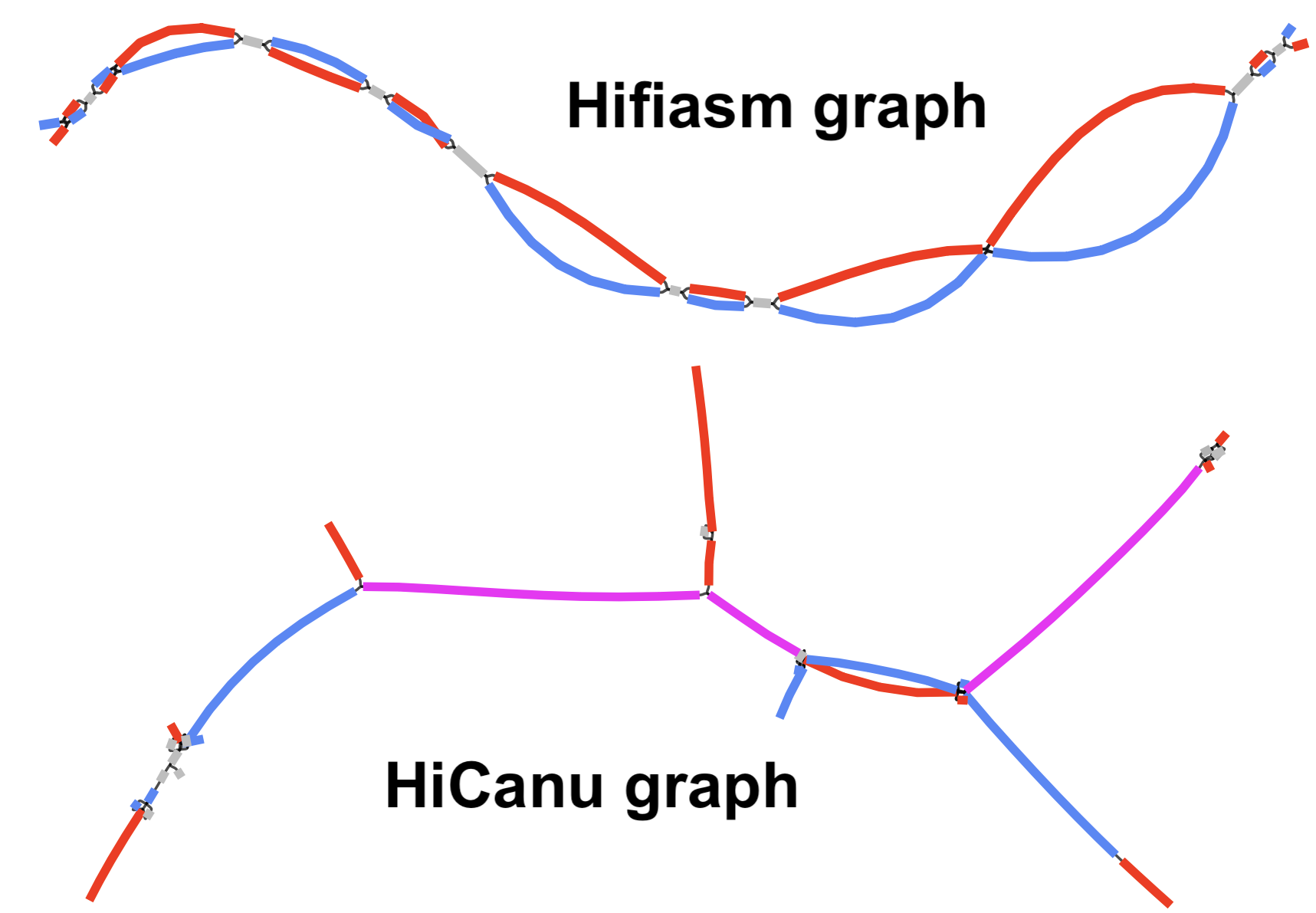}
\caption{{\bf Example hifiasm and HiCanu assembly graphs.} The graphs were
generated from HG002 reads mapped to chr11:19,310,012-21,493,943. Red bars correspond
to unitigs matching the maternal haplotype, blue to paternal, grey to homozygous
unitigs present on both parental haplotypes, and pink bars correspond to wrongly phased unitigs that join paternal and
maternal haplotypes.} \label{fig3}
\end{figure}

It may be tempted to migrate the hifiasm graph binning algorithm to HiCanu. In
practice, however, initially designed to produce best overlap graphs, HiCanu
often breaks bubbles in the string graph and misjoins unitigs from different
parental haplotypes (Fig.~\ref{fig3}). Implementing graph trio binning on
top of HiCanu would lead to fragmented assemblies at these broken bubbles.
Graph binning is a unique feature of hifiasm.

\begin{table}[!tb]
\captionsetup{singlelinecheck = false, justification=justified}
\footnotesize
\caption{Statistics of haplotype-resolved human assemblies}
{
\begin{tabular*}{\textwidth}{@{\extracolsep{\fill}} clcccccrrc}
\hline
\multirow{2}{*}{Dataset} & \multirow{2}{*}{Assembly} & \multirow{2}{*}{QV} & \makecell[c]{NG50} & \makecell[c]{Resolved} & \makecell[c]{Switch}      & \makecell[c]{Hamming} & \makecell[c]{FNR} & \makecell[c]{FDR} & \makecell[c]{Duplicated} \\
                         &                           &                     & (Mb)               & BACs (\%)              & \makecell[c]{error (\%)}  & error (\%)            & (\%)              & (\%)              & genes (\%) \\

\hline
\multirow{5}{*}{\makecell{HG00733}}
& hifiasm (trio)         & 49.8 & 35.1 & 95.5 & 0.09 & 0.26 & 2.74  &     & 0.36 \\
& HiCanu (trio)          & 47.3 & 10.2 & 92.7 & 0.04 & 0.04 & 6.18  &     & 1.78 \\
& Peregrine (trio)       & 42.2 & 19.1 & 39.7 & 0.10 & 0.23 & 12.34 &     & 0.29 \\
& Peregrine (Strand-seq) & 45.8 & 26.6 & 46.9 & 0.18 & 0.72 & 3.99  &     & 0.15 \\

\hline
\multirow{3}{*}{\makecell{HG002}}
& hifiasm (trio)         & 51.6 & 41.0 &      & 0.79 & 0.42 & 0.91  & 0.27 & 0.48 \\
& HiCanu (trio)          & 48.1 & 12.4 &      & 0.70 & 0.18 & 2.38  & 0.30 & 1.77 \\
& Peregrine (trio)       & 42.7 & 25.8 &      & 0.74 & 0.19 & 4.42  & 4.18 & 0.35 \\

\hline
\end{tabular*}
}
\begin{flushleft} \footnotesize{
Parental assemblies are merged together for computing QV, NG50 and BACs
resolved. Calculating NG50 assumes a diploid human genome size of 6.2 Gb.
Phased variants are called with dipcall \cite{Li:2018aa} for each pair of
parental assemblies and are compared to HG002 truth variants from GIAB
\cite{zook2019open} or HG00733 phased SNPs from HGSVC \cite{Chaisson:2019aa}.
Phasing switch error rate: percent adjacent SNP pairs that are wrongly phased.
Phasing hamming error rate: percent SNP sites that are wrongly phased. False
negative rate (FNR): percent true variants that are missed in the assembly.
False discovery rate (FDR): percent assembly-based variant calls that are not
called in the truth data. RTG's vcfeval \cite{Cleary023754} is used for
estimating variant FNR and FDR for HG002. For HG00733, FNR is estimated at
heterozygous SNP sites only; FDR is not available because HGSVC does not
provide confident regions. Percent duplicated genes measures the percentage of
single-copy genes in GRCh38 that are duplicated in the assembly, averaged
between the two parental haplotypes.}
\end{flushleft}

\label{table3}
\end{table}

\subsection*{Haplotype-resolved assembly of heterozygous human genomes}

To evaluate how well assemblers resolve both haplotypes, we applied trio binning
assembly to HG00733 and HG002. Hifiasm performs graph trio binning that
partitions a diploid assembly graph to generate the final assembly. HiCanu does
standard trio binning \cite{koren2018novo} that partitions HiFi reads upfront
and assembles the two parental partitions separately. Peregrine does not
natively support trio binning. We feed the HiCanu-partitioned reads to
Peregrine for assembly. For comparison, we also acquired a Strand-seq
HG00733 assembly \cite{porubsky2019fully} that uses the same
HiFi reads but are supplemented with additional data types for phasing.

On both datasets, trio hifiasm misses fewer variants and emits longer contigs
with higher QV and lower variant FDR than other assembly strategies
(Table~\ref{table3}). HiCanu achieves the lowest phasing error rates, but it
has the highest level of false duplications as is measured by ``duplicated
genes''. The HiCanu contig N50 is also the shortest. This is probably caused by
wrongly partitioned reads (Fig.~\ref{figf2}) in combination with HiCanu's strict
requirement of exact overlapping. Collapsing inexact overlaps, Peregrine can
remove most false duplications and is more robust to partition errors in
certain cases and can achieve longer contigs.  However, this comes with the cost of
fewer resolved BACs and increased FNR.  The Strand-seq
assembly also uses Peregrine. It is affected by false read partitions in the
same way. This assembly is not as good as hifiasm on every metric. It
is not possible to get a good all-around assembly if we perform separate
assemblies on pre-partitioned reads.

\section*{Discussion}

Hifiasm is a fast open-source \emph{de novo} assembler specifically developed
for HiFi reads. It mostly uses exact overlaps to construct the assembly graph
and can separate different alleles or different copies of a segmental
duplication involving a single segregating site. This greatly enhances its
power for resolving near identical, but not exactly identical repeats and
segmental duplications. In our evaluation, hifiasm consistently outperforms
Falcon and Peregrine which do not take the advantage of exact overlaps.

In comparison to HiCanu which is developed in parallel to our work, hifiasm is
able to generate a more complete assembly graph preserving all haplotypes
more contiguously. This enables us to implement a graph trio binning algorithm
that can produce a haplotype-resolved assembly tripling the contig N50 of a
trio HiCanu assembly. Hifiasm can generate the best haplotype-resolved human
assemblies so far.

Our graph binning algorithm can also work with reads labeled by Hi-C or
Strand-seq binning that do not require parental data. However, because existing
Hi-C or Strand-seq binning algorithms start with a collapsed assembly, they may
not work well with highly heterozygous regions not represented well in the
initial assembly. In our view, a better solution to pedigree-free phased
assembly is to map Hi-C or Strand-seq data to the hifiasm assembly graph,
group and order unitigs into chromosome-long scaffolds with the graph topology,
and then phase heterozygous events along the scaffolds. We envision that
haplotype-resolved assembly will become a common practice for both human and
diploid non-human species, though haplotype-resolved assembly may remain
challenging for polyploid plants in the near future.

\section*{Acknowledgements}
This study was supported by US National Institutes of Health (grant R01HG010040,
U01HG010971 and U41HG010972 to H.L.).

\section*{Author contributions}
H.C. and H.L. designed the algorithm, implemented hifiasm and drafted the
manuscript. H.C. benchmarked hifiasm and other assemblers. G.T.C. ran hifiasm
for \emph{S. sempervirens}, ran Peregrine for \emph{S. sempervirens} and
\emph{R. muscosa} and ran Falcon-Unzip for all datasets. H.Z. experimented the
\emph{R. muscosa} assembly with hifiasm and HiCanu. X.F. helped evaluation.

\section*{Competing interests} 
G.T.C. is an employee of Pacific Biosciences. H.L. is a consultant of
Integrated DNA Technologies, Inc and on the Scientific Advisory Boards of
Sentieon, Inc, BGI and OrigiMed. 

\section*{Methods}

\textbf{Haplotype-aware error correction.} The first step of hifiasm is the
haplotype-aware error correction. In this step, hifiasm loads all reads into
memory, and performs all-vs-all pairwise alignment between them. By utilizing the
alignment results, hifiasm is able to do haplotype phasing using heterozygous SNPs. Given a
reference read $R$ and its overlaps to related reads, we need to identify the
real SNPs and ignore the sequencing errors. To this end, hifiasm collects all
mismatches from the pairwise alignment between $R$ and its overlapping reads.  If one
mismatch is supported by three overlapping reads, hifiasm takes it as a SNP,
otherwise it would be ignored as a sequencing error (Fig. \ref{figf1}).  For a
read $Q$ overlapped with $R$, it comes from the same haplotype of $R$ only if
there is no difference on SNP sites between $Q$ and $R$.  To avoid
overcorrection and keep the informative heterozygosity variants carrying
haplotype information, hifiasm only uses the reads coming from the same
haplotype for error correction.  This strategy is also helpful to discard reads
coming from the highly repetitive regions since the base differences between these regions are
also treated as SNPs.  The sequencing errors on each read are then corrected by
the consensus method from pairwise alignment \cite{chin2013nonhybrid}. Although
in theory the pairwise-alignment-based consensus method is not as good as the
traditional Partial Order Alignment (POA) method \cite{lee2003generating}
when correcting noisy reads, our algorithm should be able to generate
comparable results with substantially less running time for HiFi reads due to
their low sequencing error rate.

The major bottleneck in this step is the all-vs-all pairwise alignment. In
order to accelerate it, existing assemblers usually first extract a spare
sample from each read, and then perform alignment on top of samples instead of
the whole sequences \cite{ruan2020fast, shafin2020nanopore}. However, the
sample-based alignment loses the details of overlaps, while both the phasing
and the error correction in hifiasm require the highly accurate base-level
alignment.  Hifiasm adopts the bit-vector algorithm \cite{myers1999fast} to
significantly reduce the alignment time.  It is able to
calculate multiple cells in alignment matrix at once using simple bit
operations, while the pairwise base-level alignment algorithms in current
assemblers need to calculate them one-by-one.  Moreover, hifiasm further
improves the alignment performance by splitting reads into relatively small
non-overlapping windows and calculating the alignment of windows.  Since each
window is small enough, we can take advantage of CPU SSE instructions to
simultaneously perform bit-vector algorithm on multiple windows
\cite{cheng2015bitmapper}.  In practice, one potential challenge of the
window-based strategy is that the alignment results on window extremities might
be not reliable.  To deal with this challenge, hifiasm re-aligns the subregions
around the boundary between two adjacent windows.

~\\
\noindent \textbf{Constructing phased assembly graphs.} After haplotype-aware
error correction, most sequencing errors have been removed while the
informative heterozygous variants are still kept.  With nearly error-free
reads, hifiasm is able to perform phasing accurately to determine if one
overlap is among the reads coming from different haplotypes (i.e. inconsistent
overlap). The next step is to build the assembly string graph
\cite{li2016minimap, myers2005fragment}. In this graph, nodes represent
oriented reads and each edge between two nodes represents the overlap
between the corresponding two reads.  Note that only consistent overlaps are
used to build the graph.  Since hifiasm builds the graph on top of nearly
error-free reads and highly accurate haplotype phasing, the produced assembly
graph of hifiasm is simpler and cleaner than those of current assemblers for
haploid genomes. However, for diploid genomes or polyploid genomes, its graph
becomes more complicated as reads from different haplotypes are clearly
separated out by phasing.  Fig. \ref{figf1} gives an example.  Since there is a
heterozygous allele on reads in orange and blue, hifiasm separates them into
two groups in which all reads in the same color belong to one group.  Only the
reads from same group are overlapped with each other.  For reads in green, they
are overlapped with the reads in both groups because the overlaps among them
are not long enough to cover at least one heterozygous allele.  As a result,
hifiasm generated a bubble in the assembly graph. Most existing assemblers aim to
produce one contiguous contig from the graph (i.e. single path in the graph) as
much as possible. They tend to collapse bubbles when
building the assembly graph. As a result, they will lose all but one allele in
each bubble. In contrast, hifiasm is designed to retain all
bubbles on the assembly graph.  Owing to the fact that there are still a few
errors at the corrected reads, hifiasm adopts a topological-aware graph
cleaning strategy to cut too short overlaps and avoid destroying substructures
embedding local phasing information like bubbles.  Hifiasm additionally records
the polyploid overlaps, which are very helpful in the following assembly
construction steps.

~\\
\noindent \textbf{Constructing a primary assembly.} The construction of the primary
assembly aims to produce contigs including one set of haplotypes but may
switch subregions between haplotypes.  In other words, each subregion in the
primary assembly only comes from one haplotype, while the corresponding
subregions of other haplotypes are removed as duplications.  In this step, most
existing assemblers follow the ``best overlap graph'' strategy or its
variants \cite{miller2008aggressive}. Their key idea is to retain longer
overlaps if there are multiple overlaps to a given read.  In contrast, hifiasm
produces a primary assembly mainly relied on the graph topological structures and
the phasing relationship among different haplotypes. Ideally, the phased
assembly graph of hifiasm should be a chain of bubbles for diploid genomes (see
Fig. \ref{fig3}). It is easy and reliable to extract primary assembly from
such chain of bubbles by bubble popping \cite{li2016minimap}. However, there
are still tips (i.e. deadend contigs broken in single end) on the assembly
graph caused by broken bubbles due to lack of coverage, phasing errors or
unresolvable repeats. To fix this problem, hifiasm proposes a three-stage
procedure. First, each bubble in the graph is reduced into a single path using
bubble popping. This step removes most duplicated subregions on different
haplotypes without hampering the contiguity of primary assembly.  Second, given
a tip unitig $T$ that is broken in one end but connected to a unitig $C$ in another
end, hifiasm checks if there are other contigs, which are also connected to
$C$, coming from the different haplotypes of $T$.  If such contigs are
identified, hifiasm removes tip $T$ so that unitig $C$ will become longer. The
reason is that for $T$, its corresponding region from different haplotype has
already been integrated into the new longer unitig $C$.  Since hifiasm records
overlaps between haplotypes (i.e. inconsistent overlaps), it can check if two contigs come from
different haplotypes. Last, hifiasm uses the ``best overlap graph''
strategy to deal with a few remaining unresolvable hard substructures on the
assembly graph. In most cases, the graph topological information and the
phasing information is more reliable than only keeping the longer overlaps. As
a result, hifiasm is able to generate a better primary assembly than current
assemblers which mainly rely on ``best overlap graph'' strategy.

~\\
\noindent \textbf{Constructing a haplotype-resolved assembly.} The phased assembly graph in hifiasm
embeds the local phasing information that is resolvable with HiFi reads.  In
this graph, the corresponding node of a homozygous read is at a single path
connecting two bubbles, while the corresponding node of a heterozygous read is
at a bubble (see Fig. \ref{figf2}). Given parental short reads, hifiasm labels
child HiFi reads with the existing k-mer based algorithm \cite{koren2018novo}.
When generating a fully phased assembly for one haplotype, hifiasm drops reads
of different haplotypes
from the graph, while using the local phasing information in graph to correct
the mispartition of global phasing. Hifiasm does not drop reads at a
single path connecting two bubbles, since these are homozygous reads that must
be contained in both haplotypes. For a bubble in which all reads are
heterozygous, hifiasm applies bubble popping to select a single best path
consisting of most reads with the expected haplotype label. If a few reads are
assigned false labels by global phasing, they are likely to be corrected
by the best path that traverses through them. In addition, instead of
dropping any read with non-expected haplotype label, hifiasm drops a contig if
the haplotype labels of most reads in it are non-expected.  

~\\
\noindent \textbf{Purging heterozygous duplications.} In the primary assembly construction step,
accurately keeping one set of haplotypes is more challenging for
haplotype-resolved assemblers.  Although the bubble popping method and the tip
removing method of hifiasm already purge large numbers of duplications from
multiple haplotypes, some duplications still remain on the primary
assembly, especially for subregions with a high heterozygosity rate.  Existing
assemblers postprocess the primary assembly using downstream tools like Purge\_dups
\cite{guan2019identifying}, which identify duplications by inexact all-vs-all
contig alignment.  If two contigs are overlapping with each other, the
overlapped regions between them are duplications. However, inexact contig
alignment might be not reliable on segmental duplications or repeats, leading
to more duplications left or overpurged repetitive regions.  To address this
duplication challenge, hifiasm re-assembles the contigs by building a string
graph regarding contigs as nodes, called a purge graph.  Given contig $A$ and
contig $B$, we define $A$ inconsistently overlaps $B$ if there are enough
reads of $A$ that are inconsistent overlapped with the reads of $B$.  Note that
hifiasm records all inconsistent overlaps among reads in the initial phased
assembly graph construction step by haplotype phasing.  In the purge graph of
hifiasm, each node is a contig, while an edge between two nodes is an inconsistent
overlap between their corresponding contigs.  Once the graph is built,
hifiasm generates the non-redundant primary assembly by simple graph cleaning.
As a result, the built-in purge duplication step of hifiasm is smoother and
more reliable than existing downstream tools.  This is because hifiasm
identifies duplications from multiple haplotypes using accurate haplotype
phasing of each read, while existing tools mainly rely on inexact contig
alignment. 

~\\
\noindent \textbf{Evaluating collapsed misassemblies for inbred samples.}
We mapped HiFi reads with minimap2 \cite{li2018minimap2} to each assembly
and then called apparent heterozygous SNPs with htsbox, a fork of samtools.
We selected biallelic SNPs such that each allele is supported by $d$ reads
where $d$ is set to 75\% of the average coverage of the sample. We then
hierarchically cluster these apparent SNPs \cite{Li:2018aa} as follows: we
merge two adjacent SNP clusters if (1) the minimum distance between them is
within 10kb and (2) the density of SNPs in the merged cluster is at least 1 per
1kb. A cluster longer than 5kb and consisting of $\ge$10 SNPs is identified as
a collapsed misassembly. Varying the thresholds changes the number of estimated
misassemblies but does not alter the relative ranking between assemblers.

~\\
\noindent \textbf{Evaluating gene completeness with asmgene.} BUSCO
\cite{simao2015busco} is a popular tool for evaluating gene completeness. It is
very helpful for new species, but is underpowered for species with high-quality
reference genomes. For example, BUSCO reports that the completeness of GRCh38
is only 94.8\%, even lower than the 95.1\% percent completeness of the CHM13
hifiasm assembly. We have also observed inconsistent BUSCO results at the gene
prediction stage. In one case, a maize gene Zm00001d004099 is mapped perfectly
(without mismatches or gaps) to both the Peregrine and hifiasm assemblies.
However, based on the 122450at3193 gene in the BUSCO catalog, BUSCO predicts
two different genes in the two assemblies and thinks hifiasm has fragmented
122450at3193. Inconsistencies like this make it difficult to compare assemblies
of similar quality.

In order to quantify completeness more accurately, we used paftools in minimap2
\cite{li2018minimap2} to calculate the
asmgene scores. Unlike BUSCO, asmgene scores are generated using reference
genome. It first aligns the cDNAs from EnsEMBL (v99 for human and mouse and
plant v47 for maize) to both reference genome and assembly by
minimap2, and then select single-copy genes in the reference genome. After that, it
compares the alignment results of these genes in the reference genome to those in
the assembly.  In our experiments, it provided accurate completeness evaluation for
\emph{M. musculus}, \emph{Z. mays} and human since these species have
high-quality reference genomes.

We also used asmgene to measure the resolution of genes in segmental
duplicates. Similarly, the first
step is to align the cDNAs to both reference genome and assembly by minimap2.
It then selects potential genes in segmental duplications that are aligned to
multiple regions of the
reference genome, and checks how many these genes were kept multiple times on
assembly.

\section*{Data availability} 

All HiFi data were obtained from NCBI Sequence Read Archive (SRA): SRR11606869
for \emph{Z. mays}, SRR11606870 for \emph{M. musculus}, SRR11606867 for
\emph{F.  $\times$ ananassa}, SRR11606868 and SRR12048570 for \emph{R.  muscosa}, SRP251156 for
\emph{S.  sempervirens}, SRR11292120 through SRR11292123 for CHM13, ERX3831682
for HG00733, and four runs (SRR10382244, SRR10382245, SRR10382248 and
SRR10382249) for HG002. For trio binning and computing QV, short reads were
also downloaded: SRR7782677 for HG00733, ERR3241754 for HG00731 (father),
ERR3241755 for HG00732 (mother) and SRX1082031 for CHM13.  GIAB's
``Homogeneity Run01'' short-read runs were used for the HG002 trio.  These
HG002 reads were downsampled to 30-fold coverage. All hifiasm assemblies
produced in this work are available at
\href{ftp://ftp.dfci.harvard.edu/pub/hli/hifiasm/submission/}{ftp://ftp.dfci.harvard.edu/pub/hli/hifiasm/submission/}.

\section*{Code availability} 
Hifiasm is available at
\href{https://github.com/chhylp123/hifiasm}{https://github.com/chhylp123/hifiasm}.

\bibliography{hifiasm}

\end{document}


\maketitle

\section{Software commands}

\subsection{Hifiasm}\label{sec:hifiasm}
To produce primary assemblies of homozygous samples
(\textit{M. musculus}, \textit{Z. mays} and CHM13), hifiasm (version 0.7) was
run with the following command which does not purge haplotig duplications:
\begin{quote}
\footnotesize\tt hifiasm -o \symbol{60}outputPrefix\symbol{62} -t \symbol{60}nThreads\symbol{62} -l0 \symbol{60}HiFi-reads.fasta\symbol{62}
\end{quote}
For heterozygous samples, hifiasm was run with the following command:
\begin{quote}
\footnotesize\tt hifiasm -o \symbol{60}outputPrefix\symbol{62} -t \symbol{60}nThreads\symbol{62} \symbol{60}HiFi-reads.fasta\symbol{62}
\end{quote}
We added `\texttt{-D10}' for the octoploid \textit{F. $\times$ ananassa}
because the default k-mer cutoff seems too low:
\begin{quote}
\footnotesize\tt hifiasm -o \symbol{60}outputPrefix\symbol{62} -t \symbol{60}nThreads\symbol{62} -D10 \symbol{60}HiFi-reads.fasta\symbol{62}
\end{quote}
For trio-binning assembly, we first built the paternal trio index and the
maternal trio index by yak (version r55) with the following commands:   
\begin{quote}
\footnotesize\tt yak count -b37 -t \symbol{60}nThreads\symbol{62} -o \symbol{60}pat.yak\symbol{62} \symbol{60}paternal-short-reads.fastq\symbol{62}\\
\footnotesize\tt yak count -b37 -t \symbol{60}nThreads\symbol{62} -o \symbol{60}mat.yak\symbol{62} \symbol{60}maternal-short-reads.fastq\symbol{62}
\end{quote}
and then we produced the paternal assembly and the maternal assembly with the
following command:   
\begin{quote}
\footnotesize\tt hifiasm -o \symbol{60}outputPrefix\symbol{62} -t \symbol{60}nThreads\symbol{62} -1 \symbol{60}pat.yak\symbol{62} -2 \symbol{60}mat.yak\symbol{62} \symbol{60}HiFi-reads.fasta\symbol{62}
\end{quote}

\subsection{Falcon-Unzip}
Falcon-kit (version 1.8.1) was run with the following HiFi-specific options:   
\begin{quote}
\footnotesize\tt
length\_cutoff\_pr = 8000 \\
ovlp\_daligner\_option = -k24 -h1024 -e.98 -l1500 -s100 \\
ovlp\_HPCdaligner\_option = -v -B128 -M24 \\
ovlp\_DBsplit\_option = -s400 \\
overlap\_filtering\_setting = --max-diff 200 --max-cov 200 --min-cov 2 --n-core 24 --min-idt 98 --ignore-indels
\end{quote}
\noindent Falcon-unzip-kit (version 1.3.7) was run with default options.

\subsection{HiCanu}
For primary assembly, HiCanu (version 2.0) was run with the following command
line:
\begin{quotation}
\footnotesize\tt\noindent
canu -p asm -d \symbol{60}outDir\symbol{62} genomeSize=\symbol{60}GSize\symbol{62} useGrid=false maxThreads=\symbol{60}nThreads\symbol{62} \symbol{92} \\
\indent -pacbio-hifi \symbol{60}HiFi-reads.fasta\symbol{62}
\end{quotation}
The contigs labeled by `\texttt{suggestedBubbles=yes}' were removed from the
primary assembly. For trio-binning assembly, we ran HiCanu in two steps as
recommended. We partitioned the HiFi reads by parental short reads with
the following command:   
\begin{quotation}
\footnotesize\tt\noindent
canu -haplotype -p asm -d \symbol{60}outDir\symbol{62} genomeSize=\symbol{60}GSize\symbol{62} useGrid=false \symbol{92} \\
\indent maxThreads=\symbol{60}nThreads\symbol{62} -haplotypePat \symbol{60}pat-reads.fq\symbol{62} -haplotypeMat \symbol{60}mat-reads.fq\symbol{62} \symbol{92} \\
\indent -pacbio-raw \symbol{60}HiFi-reads.fasta\symbol{62}
\end{quotation}
Note that `\texttt{-pacbio-raw}' was used to partition HiFi reads followed the
document of HiCanu. We then perform HiCanu assemblies on partitioned reads.

\subsection{Peregrine}
For primary assembly, Peregrine (version 0.1.6.1) was run with the following
command, where 48 is the number of threads in use:
\begin{quotation}
\footnotesize\tt\noindent
docker run -it -v \symbol{60}workDir\symbol{62}:/wd --user \$(id -u):\$(id -g) cschin/peregrine:0.1.6.1 asm \symbol{92} \\
\indent /wd/Input.fnfo 48 48 48 48 48 48 48 48 48 --with-consensus --with-alt --shimmer-r 3 \symbol{92} \\
\indent --best\_n\_ovlp 8 --output \symbol{60}outDir\symbol{62}
\end{quotation}
For trio-binning assembly, we first used HiCanu to partition HiFi reads by
parental short reads, and then assembled the each haplotype individually by
Peregrine.

\subsection{Purge\_dups}
Purge\_dups (version 1.2.3) was used to postprocess the output primary
assemblies of HiCanu for all heterozygous samples. The commands are as follows:   
\begin{quote}
\footnotesize\tt
minimap2 -I6G -xmap-pb \symbol{60}contigs.fa\symbol{62} \symbol{60}HiFi-reads.fasta\symbol{62} -t \symbol{60}nThreads\symbol{62} \symbol{62} \symbol{60}read-aln.paf\symbol{62} \\
bin/pbcstat \symbol{60}read-aln.paf\symbol{62} \\
bin/calcuts PB.stat \symbol{62} cutoffs \\
bin/split\_fa \symbol{60}contigs.fa\symbol{62} \symbol{62} \symbol{60}split.fa\symbol{62} \\
minimap2 -I6G -xasm5 -DP \symbol{60}split.fa\symbol{62} \symbol{60}split.fa\symbol{62} -t \symbol{60}nThreads\symbol{62} \symbol{62} \symbol{60}ctg-aln.paf\symbol{62} \\
bin/purge\_dups -2 -T cutoffs -c PB.base.cov \symbol{60}ctg-aln.paf\symbol{62} \symbol{62} \symbol{60}dups.bed\symbol{62} \\
bin/get\_seqs \symbol{60}dups.bed\symbol{62} \symbol{60}contigs.fa\symbol{62}
\end{quote}
Since running Purge\_Dups in default cannot produce primary assembly of HiCanu
with right size for HG002, we manually adjusted the cutoffs thresholds of
Purge\_Dups as follows ``{\tt 5 7 11 30 22 42}''.

\subsection{Running asmgene}
We aligned the cDNAs to the reference genome and contigs by minimap2 r974 and
evaluated the gene completeness with paftools.js from the minimap2 package:
\begin{quote}
\tt\footnotesize minimap2 -cxsplice:hq -t \symbol{60}nThreads\symbol{62} \symbol{60}contigs.fa\symbol{62} \symbol{60}cDNAs.fa\symbol{62} \symbol{62} \symbol{60}aln.paf\symbol{62} \\
\tt\footnotesize paftools.js asmgene -i.97 \symbol{60}ref.paf\symbol{62} \symbol{60}asm.paf\symbol{62}
\end{quote}
We set the sequence identity threshold to be 97\% with `\texttt{-i.97}' to
tolerate low per-base accuracy of ONT assemblies. For trio binning assemblies,
we added option `\texttt{-a}' to evaluate genes mapped to the autosomes only.
When evaluating multi-copy genes retained in an assembly, we replaced `{\tt
-i.97}' to `{\tt -i.99}' to increase the resolution.

\subsection{Computing NGA50}
We used minigraph (version 0.10-dirty-r361) and paftools (version
2.17-r974-dirty) to calculate the NGA50 of each asssembly:
\begin{quote}
\tt\footnotesize
minigraph -xasm -K1.9g --show-unmap=yes -t \symbol{60}nThreads\symbol{62} \symbol{60}ref.fa\symbol{62} \symbol{60}asm.fa\symbol{62} \symbol{62} \symbol{60}asm.paf\symbol{62} \\
paftools.js asmstat \symbol{60}ref.fa.fai\symbol{62} \symbol{60}asm.paf\symbol{62}
\end{quote}
In comparison to minimap2, minigraph tends to generate longer alignments and is
more robust to highly variable regions.

\subsection{BUSCO}
BUSCO (version 3.0.2) was used with the following command:   
\begin{quote}
\tt\footnotesize python3 run\_BUSCO.py -i \symbol{60}asm.fa\symbol{62} -m genome -o \symbol{60}outDir\symbol{62} -c \symbol{60}nThreads\symbol{62} -l \symbol{60}lineage\_dataset\symbol{62}
\end{quote}
where `\texttt{lineage\_dataset}' is set to \emph{tetrapoda} for \emph{R.
muscosa} and set to \emph{embryophyta} for \emph{F. $\times$ ananassa} and
\emph{S. sempervirens}.

\subsection{Determining resolved BACs}
The resolution of BAC for different assemblies was evaluated using the
pipeline at: \url{https://github.com/skoren/bacValidation}, except that we
added option `{\tt -I6g}' to minimap2. The BAC libraries
for CHM13 and HG00733 can be found at
\url{https://www.ncbi.nlm.nih.gov/nuccore/?term=VMRC59+and+complete} and
\url{https://www.ncbi.nlm.nih.gov/nuccore/?term=VMRC62+and+complete},
respectively.

\subsection{Running yak evaluation}
We used yak (version r55) to measure the per-base consensus accuracy (QV), the
switch error rate and the hamming error rate. For QV evaluation, we first built
the index for the short reads coming from the same sample: 
\begin{quote}
\tt\footnotesize
yak count -b37 -t \symbol{60}nThreads\symbol{62} -o \symbol{60}sr.yak\symbol{62} \symbol{60}short-reads.fastq\symbol{62} \\
yak qv -t \symbol{60}nThreads\symbol{62} \symbol{60}sr.yak\symbol{62} \symbol{60}contigs.fa\symbol{62}
\end{quote}
To evaluate the switch error rate and the hamming error rate, we first built
the indexes from the paternal short reads as in section~\ref{sec:hifiasm} and
then estimate k-mer based error rates as follows:
\begin{quote}
\tt\footnotesize yak trioeval -t \symbol{60}nThreads\symbol{62} \symbol{60}pak.yak\symbol{62} \symbol{60}mat.yak\symbol{62} \symbol{60}contigs.fa\symbol{62}
\end{quote}

\subsection{Dipcall}
For the male sample HG002, we ran dipcall (version 0.1) as follows: 
\begin{quotation}
\tt\footnotesize\noindent
dipcall.kit/run-dipcall -x dipcall.kit/hs37d5.PAR.bed \symbol{60}prefix\symbol{62} hs37d5.fa \symbol{92}\\
\indent\symbol{60}pat-asm.fa\symbol{62} \symbol{60}mat-asm.fa\symbol{62} \symbol{62} \symbol{60}prefix.mak\symbol{62} \\
make -j2 -f \symbol{60}prefix.mak\symbol{62}
\end{quotation}
For the female sample HG00733, we removed option `\texttt{-x}'.
We used the GRCh37 variant of `\texttt{hs37d5.fa}' here because GIAB works best
with hs37d5.

\subsection{Evaluating collapsed misassemblies for inbred samples}
We used scripts at:
\url{https://github.com/lh3/CHM-eval/blob/master/misc/clustreg.js}, and
\url{https://github.com/lh3/CHM-eval/blob/master/misc/select-collapse-het.js}.
The commands are as follows: 
\begin{quotation}
\tt\footnotesize\noindent
minimap2 -axasm20 -t \symbol{60}contigs.fa\symbol{62} \symbol{60}HiFi-reads.fasta\symbol{62} \symbol{124} samtools sort -o \symbol{60}aln.bam\symbol{62} -\\
htsbox pileup -vcf \symbol{60}contigs.fa\symbol{62} -q20 -Q20 -l5000 -S5000 -s5 \symbol{60}aln.bam\symbol{62} \symbol{62} \symbol{60}var.vcf\symbol{62}\\
./select-collapse-het.js -c \symbol{60}readCoverage\symbol{62} \symbol{60}var.vcf\symbol{62} $|$ ./clustreg.js -n10
\end{quotation}
where `\texttt{-l}' and `\texttt{-S}' filter out alignments short than 5kb.

\bibliography{hifiasm}      